\documentclass[]{pasj02} 
\usepackage[switch,mathlines]{lineno} 
\usepackage{natbib} 

\jyear{2025}
\Received{2025/07/03}
\Accepted{2025/10/08}

\newcommand{\Ni}{\ensuremath{^{56}\mathrm{Ni}}\xspace}
\newcommand{\Co}{\ensuremath{^{56}\mathrm{Co}}\xspace}

\newcommand{\Msun}{\,\ensuremath{\mathrm{M}_\odot}\xspace}
\newcommand{\Rsun}{\,\ensuremath{\mathrm{R}_\odot}\xspace}

\newcommand{\kmps}{\,\ensuremath{\mathrm{km~s^{-1}}}\xspace}

\graphicspath{{./}{figures/}} 

\usepackage{xspace}

\begin{document} 

\title{Type Ibn supernovae from ultra-stripped supernova progenitors}

\author{
 Takashi J. \textsc{Moriya},\altaffilmark{1,2,3}\altemailmark\orcid{0000-0003-1169-1954} \email{takashi.moriya@nao.ac.jp}
 Bernhard \textsc{M\"uller},\altaffilmark{3}\orcid{0000-0002-4470-1277}
 Sergei I. \textsc{Blinnikov},\altaffilmark{4,5,6}\orcid{0000-0002-5726-538X}
 Marina \textsc{Ushakova},\altaffilmark{5,6}
 Elena I. \textsc{Sorokina},\altaffilmark{5,6}
 Thomas M. \textsc{Tauris},\altaffilmark{7}\orcid{0000-0002-3865-726}
 and
 Alexander \textsc{Heger}\altaffilmark{3,8}\orcid{0000-0002-3684-1325}
}
\altaffiltext{1}{National Astronomical Observatory of Japan, National Institutes of Natural Sciences, 2-21-1 Osawa, Mitaka, Tokyo 181-8588, Japan}
\altaffiltext{2}{Graduate Institute for Advanced Studies, SOKENDAI, 2-21-1 Osawa, Mitaka, Tokyo 181-8588, Japan}
\altaffiltext{3}{School of Physics and Astronomy, Monash University, Clayton, VIC 3800, Australia}
\altaffiltext{4}{Kavli Institute for the Physics and Mathematics of the Universe, The University of Tokyo, Kashiwa, 277-8583 (Kavli IPMU, WPI) Japan}
\altaffiltext{5}{NRC Kurchatov Institute 123182 Moscow, Russia}
\altaffiltext{6}{Sternberg Astronomical Institute, Lomonosov Moscow State University, Moscow, 119234 Russia}
\altaffiltext{7}{Department of Materials and Production, Aalborg University, Skjernvej 4A, DK-9220 Aalborg Øst, Denmark}
\altaffiltext{8}{Argelander-Institut f{\"u}r Astronomie, University of Bonn, Auf dem H{\"u}gel 71, 53121 Bonn, Germany}




\maketitle

\begin{abstract}
Ultra-stripped supernovae are core-collapse supernovae from progenitors that lose a significant fraction of mass because of the binary interactions with their compact companion stars. Ultra-stripped supernovae have been connected to fast-evolving faint Type~Ib or Ic supernovae. Here, we show that in some cases ultra-stripped supernovae can result in Type~Ibn supernovae. Progenitors of ultra-stripped supernovae may trigger violent silicon burning shortly before the core collapse, leading to mass ejection that results in a dense circumstellar matter. By taking an ultra-stripped supernova progenitor that loses $0.2~\Msun$ at 78~days before the core collapse, we compute the light-curve evolution of the ultra-stripped supernova within the dense circumstellar matter. The core collapse results in a supernova explosion with an ejecta mass of $0.06~\Msun$ and an explosion energy of $9\times 10^{49}~\mathrm{erg}$. Because the dense circumstellar matter is more massive than the supernova ejecta, the ejecta are immediately decelerated and the light curve is powered mainly by the circumstellar interaction. Therefore, this ultra-stripped supernova is likely observed as a Type~Ibn supernova. We suggest that some Type~Ibn supernovae may originate from ultra-stripped supernova progenitors losing significant mass shortly before their explosion due to violent silicon burning.
\end{abstract}


\section{Introduction}
Ultra-stripped supernovae (SNe) are explosions of massive stars that lose a significant fraction of their envelopes before their explosion \citep{tauris2013,tauris2015,jiang2021}. The significant mass stripping is caused by the binary interaction with their compact companion star --- a neutron star, black hole, or massive white dwarf --- during so-called Case~BB Roche-lobe overflow (RLO) from a naked helium-star, following common envelope and spiral-in evolution \citep{tvdh23}.  Ultra-stripped SNe are considered to be the major evolutionary pathway to form double neutron star systems, including the coalescing binary neutron stars observed in gravitational waves \citep[e.g.,][]{tauris2017}.
Ultra-stripped SNe have also been connected with the formation of young pulsars in tight binaries with a massive first-born white dwarf companion \citep{vbv+20}. A general condition for ultra-stripped SNe is their association with the formation of the second-born neutron star in a binary. This stems from the fact that envelope stripping by a non-degenerate companion is comparatively inefficient \citep{ywl10}. \citet{rpe+23} suggested that the neutron star in the Be-star/X-ray binary SGR~0755$-$2933 (i.e., the first-born compact object) was created in an ultra-stripped SN. However, this conclusion was based on an erroneous analysis of the kinematic effects of the SN on the surviving binary \citep{llp+24}. For the possibility of ultra-stripped SNe in black hole binaries, see, e.g., \citet{jiang2023}. 

Identifying ultra-stripped SNe in transient surveys is important to constrain their nature and event rates so that we can test our understanding of binary stellar evolution and the formation pathways of gravitational wave sources \citep[e.g.,][]{kruckow2018,hijikawa2019,wei2024}.
The ejecta masses of ultra-stripped SNe are suggested to be less than around $0.2~\Msun$, although they may reach higher values ($0.2-0.5~\Msun$) in cases where the pre-SN mass stripping is less efficient \citep[depending on the orbital period and the initial helium-star mass prior to Case~BB RLO,][]{tauris2015}. Ultra-stripped SNe can be of spectral Type~Ib or Ic, depending on the amount of residual helium in the envelope of the collapsing star. The current paradigm is that ultra-stripped SNe have small explosion energies of the order of $10^{50}~\mathrm{erg}$ \citep{suwa2015,muller2018,muller2019} with a small amount of \Ni production ($\sim 0.01~\Msun$, \citealt{suwa2015,yoshida2017,muller2018,muller2019,sawada2022}).  As a result, current models predict that ultra-stripped SNe produce rapidly-evolving faint Type~Ib and Ic SNe (e.g., \citealt{tauris2013,suwa2015,moriya2017,maunder2024,maunder2025}, but see also \citealt{hotokezaka2017,sawada2022,mor2023} for the possibilities to result in luminous SNe). Some observed SNe have properties consistent with predictions for ultra-stripped SNe and have hence been associated with ultra-stripped SNe \citep[e.g.,][]{tauris2013,drout2013,moriya2017,de2018,ertini2023,shivkumar2023,agudo2023,yan2023,das2024,moore2025}.

Interestingly, some observed ultra-stripped SN candidates show signatures of the existence of dense circumstellar matter (CSM). For example, SN~2014ft (iPTF14gqr) shows early light-curve decline and narrow helium emission features within two days after explosion \citep{de2018}. These features are consistent with having a confined dense CSM with $0.008~\Msun$ extending to $3\times 10^{13}~\mathrm{cm}$. \citet{de2018} suggested that the ultra-stripped SN progenitor experienced strong mass loss shortly before its explosion.

In previous work of the ultra-stripped SN progenitor evolution and explosion, \citet{muller2018} found that their $2.8~\Msun$ helium star progenitor model experiences a significant mass loss at $78$~days before core collapse due to violent silicon burning \citep[e.g.,][]{woosley2015,woosley2019}. This violent silicon burning initiates a strong sound wave traveling towards the stellar surface and results in the ejection of $0.2~\Msun$. The core-collapse simulation of \citet{muller2018} found that this model resulted in an ultra-stripped SN with an ejecta mass of $0.06~\Msun$. \citet{maunder2024} investigated the observational properties of this ultra-stripped SN, but they did not take into account the presence of the CSM formed by the mass ejection. The observational properties of the ultra-stripped SN can be significantly affected by the CSM, because the CSM mass ($0.2~\Msun$) is about three times more massive than the ejecta mass ($0.06~\Msun$) in this model. In the present work, we investigate the effect of the CSM on the observational properties of the ultra-stripped SN.

This paper is organized as follows. We first introduce the modified \texttt{STELLA} code used in this work in Section~\ref{sec:stella}. We introduce the ultra-stripped SN progenitor and the initial conditions for our synthetic light-curve calculations in Section~\ref{sec:progenitor}. We present our synthetic light curves in Section~\ref{sec:lightcurve} and discuss the results in Section~\ref{sec:discussion}. Our conclusions are given in Section~\ref{sec:conclusions}.

\section{The \texttt{STELLA} code}\label{sec:stella}
In this section, we describe the one-dimensional multi-frequency radiation hydrodynamics code \texttt{STELLA} \citep{blinnikov1998,blinnikov2000,blinnikov2006}, which we use to obtain synthetic light curves in this paper. \texttt{STELLA} implicitly solves the time-dependent equations for the first two angular moments of radiation intensity averaged over a frequency bin by using the variable Eddington factor method. In this paper, we use the standard 100 frequency bins uniformly spaced on a log scale between $6.3\times 10^{13}~\mathrm{Hz}$ ($ \lambda = 5\times 10^4$~\AA) and $2.8\times 10^{8}~\mathrm{Hz}$ ($ \lambda = 1$~\AA). Spectral energy distributions (SEDs) with this frequency grid are calculated at each timestep. Multicolor light curves with any filters can be evaluated by convolving the filter functions with the synthetic SEDs. \texttt{STELLA} assumes spherical symmetry, and the mass ejection by the shell flash is likely to result in the formation of a roughly spherically symmetric CSM. Thus, \texttt{STELLA} is expected to provide reasonable predictions for the light-curve evolution of the system we are interested in. We made some modifications in the \texttt{STELLA} code in this study as described below.

\subsection{Modifications in standard \texttt{STELLA} algorithm for error controls}

\texttt{STELLA} employs the method of lines to solve the system of partial differential equations (PDE) for radiation hydrodynamics thus transforming them into a huge system of ordinary differential equations (ODE).
Standard \texttt{STELLA} runs \citet{Gear1971} or \citet{Brayton1972} integrator for numerical solution of the ODE system thus obtained.

Predictor-corrector schemes, widely used for solving stiff differential equations in simulations, typically estimate numerical errors for each timestep as the sum of squared deviations between corrector and predictor across all variables, that is, they rely on the Euclidean norm of the error vector. 
However, this approach risks masking localized inaccuracies critical to the physical correctness of the model. 
Instead of aggregating and averaging errors over all variables in the model, 
the new modifications in the \texttt{STELLA} code that are implemented in this study define the error used for the control of the next timestep and order of the integration method, as the maximum individual deviation observed among all variables. 
This change ensures that the prediction of no single variable deviates excessively from the true value, prioritizing the accuracy of the worst case.

\subsection{Modifications for tuning of smearing factors}

When SN ejecta collide with a dense CSM, the shocked gas cools rapidly by radiation in the optically thin regime. 
This cooling leads to the formation of a thin dense shell between the ejecta and the CSM. 
In one-dimensional models, this cooling becomes catastrophic: the shell gets denser, making cooling even more efficient and leading to unrealistically thin structures.

These shells, however, are hydrodynamically unstable to non-radially symmetric perturbations. 
Multi-dimensional effects such as Rayleigh-Taylor instabilities (seen in multi-dimensional simulations, e.g., \citealp{2022Urvachev,suzuki2019}) lead to bending and smearing of the shell.
A large fraction of the kinetic energy of radial motions is transformed into lateral motions.
This reduces the amount of kinetic energy that dissipates into radiation. 
The effective shell thickness must be much larger than found in one-dimensional simulations.
This makes radiative cooling less efficient in multi-dimension than in one dimension.

To mimic these multi-dimensional effects in one-dimensional codes like \texttt{STELLA}, a ``smearing term'' is added to the equations.
The standard \texttt{STELLA} treatment of the smearing is described in \citet{blinnikov1998} and \citet{moriya2013gy}.
We repeat this description here giving more details and pointing out the current modifications.
This modified version of the smearing term described below was first implemented in \texttt{STELLA} calculations in \cite{2016ApJ...829...17S}.

The Euler equation in Lagrangian coordinates for a spherically-symmetric one-dimensional flow with radial velocity $u$, is 
\begin{equation}
\frac{\partial u}{\partial t} = - 4\pi
r^2\frac{\partial (P+Q)}{\partial m}+g+a+b\;.
\label{accel}
\end{equation}
Here, $P=P(\rho,T)$ is the material pressure, $\rho$ is the density, $Q$
is an artificial viscous pressure, $m$ is the Lagrangian coordinate (the mass inside a spherical shell with radius $r$),
\begin{equation}
g = - \frac{Gm}{r^2}\;,
\end{equation}
$G$ is the Newton's gravitational constant, 
$a$ is the acceleration
due to the radiation momentum flux,  and 
$b$ is an additional radial acceleration used for the smearing of large density contrasts developing in one-dimensional modeling of shock propagation in radiating fluids.
Note that \citet{blinnikov1998} use the notation $a_{\rm mix}$ instead of $b$.

To find the acceleration term $b_i$ used for smearing at the mesh point
number $i$ in the current version of \texttt{STELLA} we first introduce a factor
 \( \psi_i \) defined as 
\begin{equation}
\begin{split}
    \psi_i = B_\mathrm{q} \frac{dm_i} {1 + \tau_i/\tau_{\max}} \left( \frac{r_i}{r_{i}-r_{i-1}}\right)^{\!\!N_{\rm RT}-1}\\
             \times   \left( -\min(0, {\rm div}_\mathrm{d}\, {u}|_{i-1/2} - \epsilon_\mathrm{q})  \right) .
\label{psi}
\end{split}
\end{equation}
Here $\hbox{\rm div}_\mathrm{d}$ is a difference approximation to the exact divergence of velocity:
\begin{align}
{\rm div}_\mathrm{d}\, {u}|_{i-1/2} &\equiv \frac{u_i r_i^2-u_{i-1}r_{i-1}^2}{r_i^2 (r_i-r_{i-1})} \nonumber \\&\approx \mathrm{div} \, {u}|_{(i-1/2)} \nonumber \\ &= \frac{1}{r^2} \frac{\partial ( ur^2)}{\partial r}|_{(i-1/2)}\;,
\end{align}
for the cell of mass $dm_i$ between radii $r_{i-1}$ and $r_{i}$.
{\bf Equation~(\ref{psi}) contains numerical parameters 
$\epsilon_\mathrm{q}$, $B_\mathrm{q}$, $\tau_{\max}$, $N_{\rm RT}$, and $D_{\rm RT}$ that can vary during calculations. We explain them below.}

A small parameter $\epsilon_\mathrm{q}$ is defined as
\begin{equation}
  \epsilon_\mathrm{q} \equiv \max(D_{\rm RT}\cdot\varepsilon|u_i|) \; \mbox{for all } \; i ,
\end{equation}
where $\varepsilon$ is the general tolerance for a timestep of \texttt{STELLA} integrator.
Introducing $\epsilon_\mathrm{q}$ allows us to take into account that we do not have
infinite accuracy, thus the smearing effect takes place also for small positive
values of the numerical divergence of velocity.

The most important parameter in Equation~(\ref{psi}) is $B_\mathrm{q}$. It can be varied over a wide range from zero up to few tens. 
Its effect was studied by~\cite{moriya2013gy}.

The parameter $N_{\rm RT}>1$ may be used to enhance the smearing in dense shells, when $r_i \to r_{i-1}$. 
In the standard setup $N_{\rm RT}=1$ and $D_{\rm RT}=1$, but other values of $N_{\rm RT}$ and $D_{\rm RT}$ may be used to enhance or reduce the effect of smearing.

The optical depth is defined crudely as that of a fully ionized gas with elements heavier than hydrogen having equal numbers of neutrons and protons:
\begin{equation}
\tau_i= \frac{0.2\,(1+X_\mathrm{H})(M-m_i)}{r_i^2} ,
\end{equation}
where $M$ is the total mass of the star, and \( m_i \) is the Lagrangian mass inside the radius \( r_i \).
Thus,  \( \tau_i \) is the optical depth of the layer between \( m_i \) and $M$.
The parameter \( \tau_{\max} \) controls the suppression of smearing at high \( \tau \).
In standard \texttt{STELLA}  version \( \tau_{\max} =30 \) was chosen.
Increasing this value allows the smearing effect to penetrate deeper into optically thick regions
(\( \tau \gg 1 \)), mimicking 3D turbulence and delaying catastrophic cooling.
In the version of \texttt{STELLA} used in the current work
\( \tau_{\max} = 10^3 \) was used.

We found that the following expression for the acceleration,
$b$, at mesh point $i$ based on $\psi_i$ from Equation~(\ref{psi}) produces satisfactory results:
\begin{equation}
b_i = \frac{2}{(dm_i+dm_{i+1})} ( \psi_i \cdot u_{i-1} - \psi_{i+1} \cdot u_{i+1}) .
\label{mixac}
\end{equation}
If the Lagrangian grid were exactly uniform then
$2dm_i/(dm_i+dm_{i+1})=2dm_{i+1}/(dm_i+dm_{i+1})\equiv 1$ and in the sum of $b_i\, u_i$
terms only the boundary ones are left.
Then the total kinetic energy is manifestly conserved
\citep[cf.][]{blinnikov1998}.

For a non-uniform grid $dm_i$ in $\psi_i$ gives more weight for more massive meshes
in the artificial dissipation although slightly violates the exact kinetic energy integral.

The Equation~(\ref{mixac}) looks very much like
a gradient of an artificial viscous pressure due to the presence of ${\rm div}\, {u}$ terms, but it is constructed in such a way,
that in effect
it does not change the kinetic energy integral.
One can easily verify this
using the relation:
\begin{equation}
u_i \,b_i = u_i\dot u_i |_b= \frac{1}{2} \left. \frac{d u_i^2}{dt} \right|_b ,
\label{termKinEn}
\end{equation}
and the
Equation~(\ref{mixac}) for $b_i$ and summing over all zone numbers $i$.
During the summation, the successive products of the terms with indices $i$ and
$i+1$ cancel each other.
The subscript $b$ indicates the part of the kinetic energy variation which is due
solely to the artificial acceleration, $b$.
Thus the term $b$ only redistributes the kinetic energy between
neighboring mass shells when there is a strong compression, i.e.
$\hbox{\rm div}\, {u} $ is negative, and, contrary to the artificial viscosity
$Q$, one does not have to include into the energy equation any
effects associated with $b$.

The factor $\psi_i$ depends on the optical depth, $\tau_i$.
It is introduced so that artificial smearing is reduced in optically thick regions
where the effect of cooling
is less efficient and multidimensional instabilities due to cooling do not develop.
The factor $\psi$ provides a normalization
of the effect of the artificial acceleration, is equal to zero
in the outermost zone (since divergence of velocity is positive there), and  kills $b$ at optical
depths $\tau \gg 1$.

These modifications allows us to compute light curves of the SN explosion with strong CSM interaction from the ultra-stripped SN progenitor described in the next section.

\section{Progenitor evolution and initial conditions}\label{sec:progenitor}

In this study we adopt the same ultra-stripped SN progenitor model that was investigated by \citet{muller2018}.  We briefly summarize its evolution until core collapse.  The progenitor is first evolved by \citet{tauris2015} from a solar-metallicity $2.8~\Msun$ helium star.  It is evolved in a binary system with a $1.34~\Msun$ neutron star.  The initial orbital period is 20~days.  This star is evolved up to the neon burning stage by \citet{tauris2015} using the \texttt{BEC} code \citep{wellstein2001}.  At this stage, the progenitor has a mass of $1.72~\Msun$ and a remaining helium envelope mass of $0.217~\Msun$.  About $0.3~\Msun$ is lost by a stellar wind and around $0.8~\Msun$ is lost by the Case~BC RLO at this moment.

The remaining evolution is computed using the \texttt{KEPLER} code \citep{weaver1978,heger2000} since the \texttt{BEC} code is not suitable for following the ensuing stellar evolution until core collapse.  The advanced burning proceeds under strong degenerate conditions due to the small core mass \citep{woosley2015,woosley2019}.  In this model, therefore, violent silicon burning is triggered at 78~days before core collapse, leading to the ejection of $0.2~\Msun$.  Most of the ejecta are composed of helium.  The progenitor mass at the time of core collapse is $1.50~\Msun$, with a remaining helium envelope mass of only $0.02~\Msun$.  This small progenitor mass makes it an ultra-stripped SN progenitor \citep{tauris2013,tauris2015}. 

\begin{figure}
 \begin{center}
  \includegraphics[width=8cm]{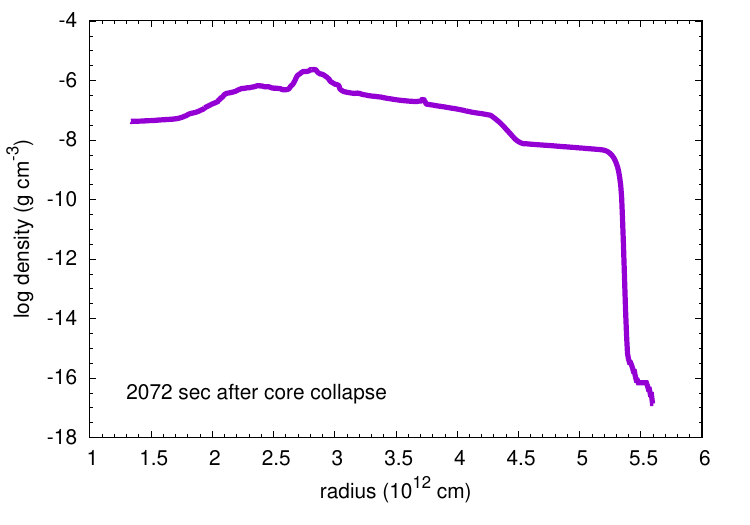}
  \includegraphics[width=8cm]{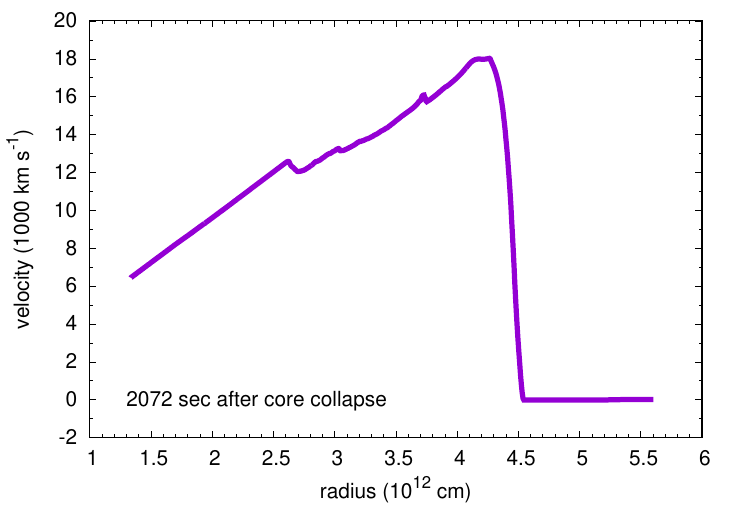}
 \end{center}
\caption{ 
Density (top) and velocity (bottom) structure of the ultra-stripped SN model shortly before the shock breakout (2,072~sec after core collapse) as obtained by \citet{muller2018}.  This structure is the initial condition for the light-curve model without CSM.
 {Alt text: Line graph. The x axis is the radius ranging from 1 to 6 in the unit of one trillion cm in both top and bottom panels. The y axis of the top panel is the density in g/cc in log ranging from -18 to -4.  The y axis of the bottom panel is the velocity in 1000~km/s ranging from -2 to 20.} 
}\label{fig:density_breakout}
\end{figure}

\begin{figure}
 \begin{center}
  \includegraphics[width=8cm]{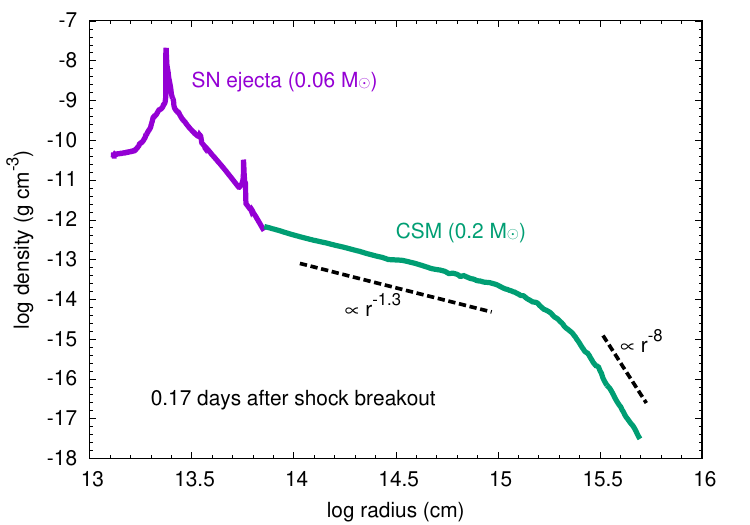}
  \includegraphics[width=8cm]{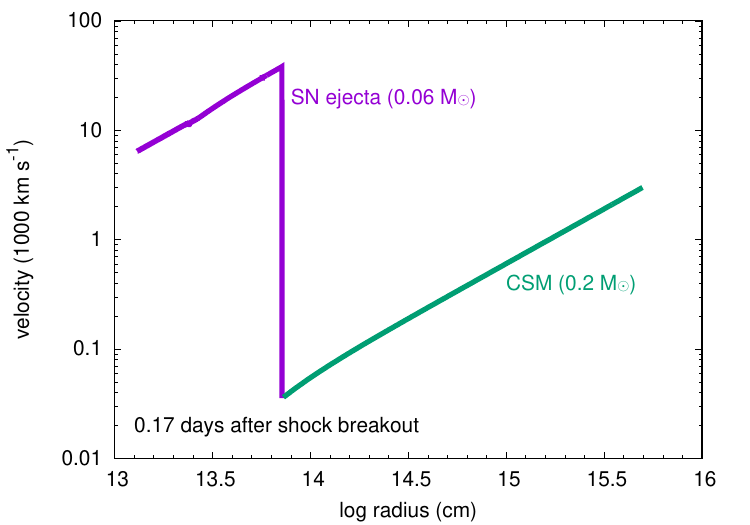}
 \end{center}
\caption{
Density (top) and velocity (bottom) structure of the CSM at 0.17~days after shock breakout. The SN ejecta at 0.17~days after shock breakout are also presented. The synthetic light curves with CSM are computed by setting this structure with the SN ejecta and CSM as the initial condition.
{Alt text: Line graph. The x axis is the radius in cm in log ranging from 13 to 16 in both top and bottom panels. The y axis of the top panel is the density in g/cc in log ranging from -18 to -7. The y axis of the bottom panel is the velocity in 1000~km/s in a log scale ranging from 0.01 to 100.} 
}\label{fig:density}
\end{figure}

\citet{muller2018} followed the subsequent core collapse and neutrino-driven explosion of this ultra-stripped SN progenitor.  First, the collapse and explosion were numerically modeled  using the relativistic neutrino radiation hydrodynamics code \texttt{CoCoNuT-FMT} \citep{muller2015}.  As in the previous study by \citet{maunder2024}, we adopt the explosion model from the two-dimensional simulation (\texttt{s2.8-2D-b}).  In this model, the core collapse and subsequent explosion were followed in two dimension until around 1~sec after the core collapse.  The explosion energy is $9\times10^{49}~\mathrm{erg}$ and the ejecta mass is $0.06~\Msun$ with the \Ni mass of $0.01~\Msun$.  They follow the explosion further using the hydrodynamics code \texttt{PROMETHEUS} \citep{fryxell1991,mueller1991} until shortly before shock breakout (2,072~sec after the core collapse). 

We spherically average and map the two-dimensional hydrodynamical and abundance profiles shortly before shock breakout to the one-dimensional \texttt{STELLA} code for the light-curve calculations.  The spherically averaged density and velocity profiles are shown in Figure~\ref{fig:density_breakout}.  The ultra-stripped SN progenitor has an extended envelope reaching out to $5.6\times 10^{12}~\mathrm{cm}$ ($81~\Rsun$) because of the violent nuclear burning that occurred 78~days before core collapse.  Thus, the shock breakout of this ultra-stripped SN progenitor occurs at around $80~\Rsun$, not at the $0.1-1~\Rsun$ that are typical radii of ultra-stripped SN progenitors \citep{tauris2015}.  The light curve without CSM is computed from this initial condition.

The $0.2~\Msun$ ejected by the violent silicon burning forms a dense CSM.  In order to obtain the CSM structure at the time of the explosion, we follow the evolution of the CSM based on the equation of motion for 78~days.  At the time of core collapse, the innermost radius of the CSM is $7.2\times 10^{13}~\mathrm{cm}$ (Figure~\ref{fig:density}).  The density of the inner CSM structure follows $\propto r^{-1.3}$ and the density of the outer CSM structure follows $\propto r^{-8}$, as expected in this kind of mass ejection \citep[e.g.,][]{tsuna2021}.  To obtain the light curves with this CSM, we first compute the ultra-stripped SN model without the CSM until the outermost layer of the SN ejecta reaches the innermost layer of the CSM at 0.17~days after the shock breakout.  At this evolution stage, we attach the structures of the SN ejecta and dense CSM to make the initial condition for the light-curve computations of the ultra-stripped SN explosion with the dense CSM.  The initial density and velocity profiles with the dense CSM are presented in Figure~\ref{fig:density}.  In this way, we avoid the numerical difficulties of including both the high-density CSM above $7.2\times 10^{13}~\mathrm{cm}$ and the low-density CSM below $7.2\times 10^{13}~\mathrm{cm}$ from the time of core collapse.  We expect that the overall light-curve evolution is not affected by our method of setting the initial condition, because it takes only 0.17~days (4.1~hrs) for the outermost layer of the SN ejecta to reach the innermost layer of the dense CSM.

\begin{figure}
 \begin{center}
  \includegraphics[width=8cm]{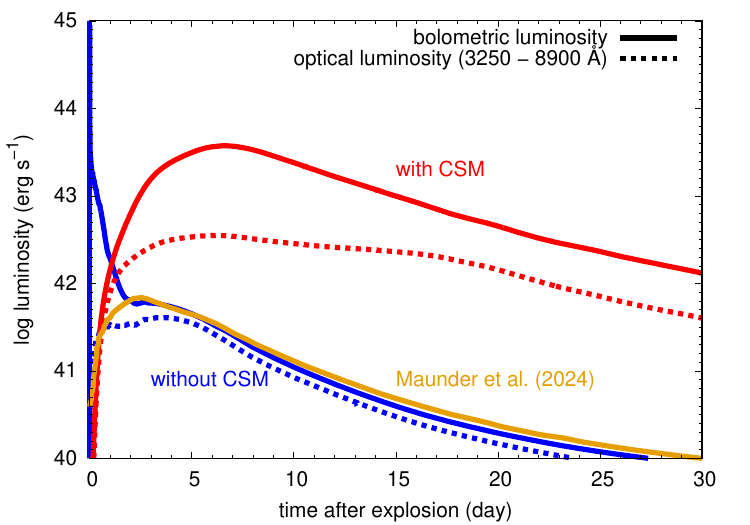} 
  \includegraphics[width=8cm]{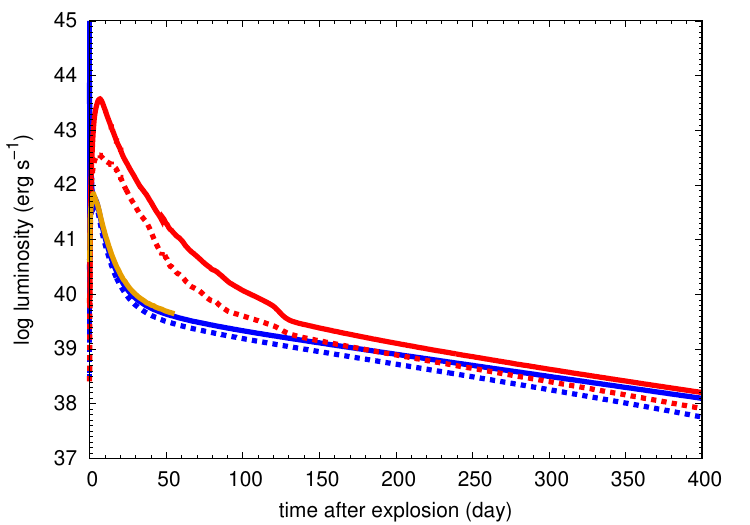} 
 \end{center}
\caption{
Top: Synthetic light curves of ultra-stripped SNe with and without CSM.  The bolometric light curves are presented by the solid lines.  The integrated luminosity in the optical wavelength range ($3250-8900$~\AA) is presented by the dashed lines.  The synthetic light-curve model from the same explosion model computed by \citet{maunder2024} is presented for comparison. Note that the light-curve model of \citet{maunder2024} does not take the initial cooling phase into account.
Bottom: The same as the top panel but with a longer time range.
 {Alt text: Line graph. The x axis is the time after explosion from 0 to 30~days in the top panel and from 0 to 400~days in the bottom panel. The y axis is the luminosity in erg/s in log ranging from 40 to 45 in the top panel and from 37 to 45 in the bottom panel.} 
}\label{fig:bolometric}
\end{figure}

\begin{figure}
 \begin{center}
  \includegraphics[width=8cm]{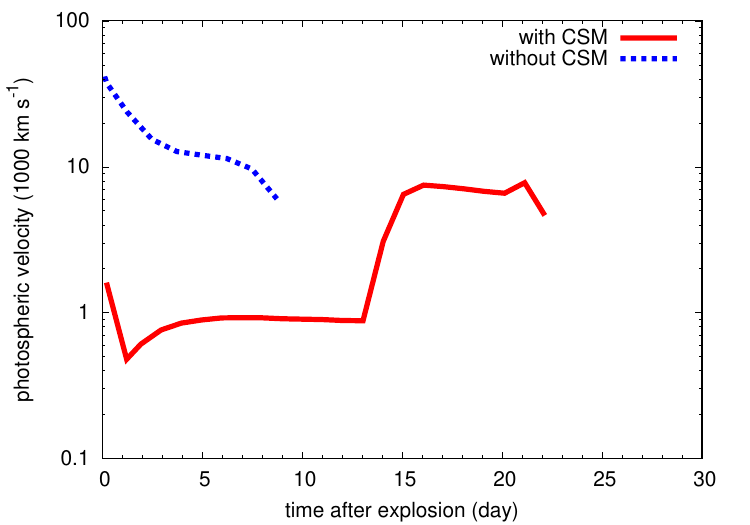} 
 \end{center}
\caption{
Photospheric velocity evolution of ultra-stripped SNe with and without CSM.  The photosphere is defined at the location where the Rosseland-mean optical depth become $2/3$.  The photospheric velocities are plotted only during the phases when the SN is optically thick.
 {Alt text: Line graph. The x axis is the time after explosion from 0 to 30~days. The y axis is the photospheric velocity in 1000~km/s in a log scale ranging from 0.1 to 100.} 
}\label{fig:photov}
\end{figure}

\begin{figure}
 \begin{center}
  \includegraphics[width=8cm]{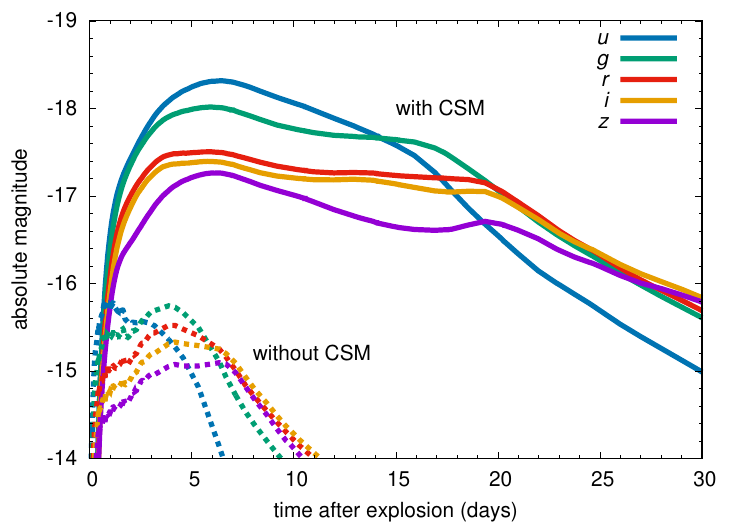} 
 \end{center}
\caption{
Synthetic light curves of ultra-stripped SNe with and without CSM in the optical ($ugriz$) bands.
 {Alt text: Line graph. The x axis is the time after explosion from 0 to 30~days. The y axis is the absolute magnitude ranging from -14 to -19. } 
}\label{fig:optical}
\end{figure}

\section{Synthetic light curves}\label{sec:lightcurve}

Figure~\ref{fig:bolometric} shows the synthetic light curves of the ultra-stripped SNe with and without the dense CSM.  We find that the CSM alters the luminosity evolution drastically.  First, we discuss the synthetic light curve without the CSM.  After shock breakout, the light curve without the CSM first shows a cooling phase for 2~days.  Ultra-stripped SN progenitors are usually compact ($0.1-1~\Rsun$) and such a long cooling phase has not been expected.  Our stripped-envelope SN progenitor, however, is expanded because of the violent nuclear burning, and the shock breakout occurs at around $80~\Rsun$.  The initial shock cooling phase shows such a slow luminosity decline because of the extended radius.  The synthetic bolometric light curve from the same progenitor in \citet{maunder2024} does not take into account the shock breakout phase, causing the discrepancy in the early light-curve behavior.  After the cooling phase, the heating from the \Ni decay starts to dominate the luminosity from around 2~days.  The luminosity evolution of our synthetic light curve and that obtained by \citet{maunder2024} are almost identical during the phase when the \Ni heating dominates the luminosity evolution.  The \Ni-powered component of the bolometric light curve rises in 2.6~days and the peak luminosity is $7\times 10^{41}~\mathrm{erg~s^{-1}}$ as found in \citet{maunder2024}.

If we add the dense CSM, the SN ejecta are immediately decelerated because the mass of the CSM ($0.2~\Msun$) exceeds the mass of the ultra-stripped SN ejecta ($0.06~\Msun$).  The bolometric light curve with the CSM rises in 6.6~days, which roughly corresponds to the diffusion time in the CSM.  The peak luminosity is $3\times 10^{43}~\mathrm{erg~s^{-1}}$, which is about 40 times more luminous than the \Ni-powered peak.  $3\times 10^{49}~\mathrm{erg}$ of radiation energy is released within 30~days after explosion, corresponding to one third of the explosion energy.  The early luminosity is dominated by the interaction-powered luminosity until around 150~days. After around 150~days, the \Co decay luminosity starts to affect the light-curve evolution.

Figure~\ref{fig:photov} shows the photospheric velocity evolution of the models with and without the CSM.  The photospheric velocity of the model without the CSM declines steadily until the entire ejecta become transparent at 9~days.  On the other hand, the photospheric velocity of the model with the CSM remains at around $900~\kmps$ until 13~days after peak when the forward shock reaches the photosphere of the dense CSM located at $8\times 10^{14}~\mathrm{cm}$. Afterwards, the photosphere remains in the cool dense shell with the photospheric velocity of $7\mathord,000~\kmps$ for about 5~days.  Then, the entire system becomes optically thin.

Figure~\ref{fig:bolometric} shows the luminosity evolution in the optical wavelength range ($3\mathord,250-8\mathord,900$~\AA).  Whereas the optical luminosity dominates in the model without the CSM around the luminosity peak, most of the luminosity in the model with the CSM is in the ultraviolet wavelength range around the bolometric luminosity peak.  Only about 10\% is in the optical wavelength range.  Figure~\ref{fig:optical} shows the \textit{ugriz}-band light curves of the models of Figure~\ref{fig:bolometric}.  The optical light curve of the model with CSM has a slow luminosity evolution and it appears to have a short plateau phase lasting for 20~days in the \textit{gri} bands, whereas the \textit{u}-band light curve has a round shape following the overall bolometric light-curve evolution.

\begin{figure}
 \begin{center}
  \includegraphics[width=8cm]{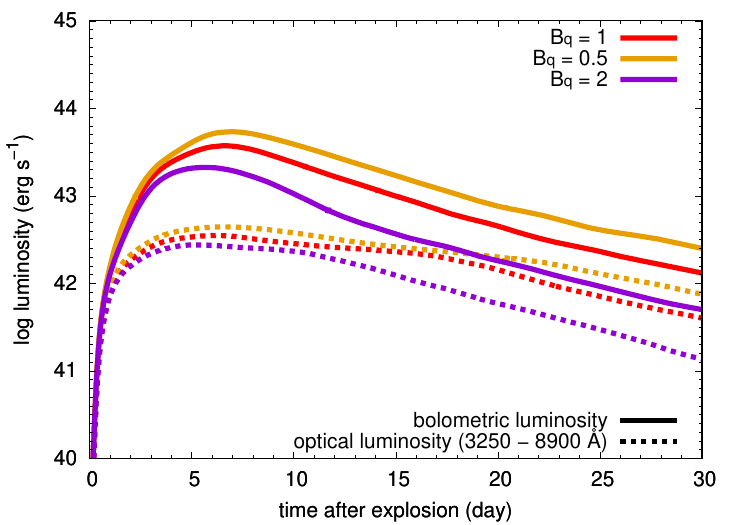} 
 \end{center}
\caption{
Synthetic light curves of the ultra-stripped SN with CSM from three different smearing factors ($B_\mathrm{q}=0.5,1,$ and $2$).  The bolometric light curves are presented by the solid lines, and the light curves in the optical wavelength range ($3250-8900$~\AA) are presented by the dashed lines.
 {Alt text: Line graph. The x axis is the time after explosion from 0 to 30~days. The y axis is the luminosity in erg/s in log ranging from 40 to 45. } 
}\label{fig:bq}
\end{figure}

We expect that the smearing factor ($B_\mathrm{q}$) in \texttt{STELLA} discussed in Section~\ref{sec:stella} affects the synthetic light curve with strong CSM interactions.  In order to quantify its effect, we computed additional light-curve models with CSM with $B_\mathrm{q} = 0.5$ and $2$.  The results are presented in Figure~\ref{fig:bq}.  As expected, the smearing factor affects the luminosity evolution with smaller $B_q$ resulting in higher luminosities. The total radiated energies within $30~\mathrm{days}$ after explosion are $5\times 10^{49}~\mathrm{erg}$ for $B_\mathrm{q}=0.5$ and $2\times 10^{49}~\mathrm{erg}$ for $B_\mathrm{q}=2$. The overall light-curve properties, however, remain unchanged.  Thus, our standard model with $B_\mathrm{q}=1$ is a good representative of the expected light curves showing the overall effects of the existence of the CSM on ultra-stripped SNe.

\section{Discussion}\label{sec:discussion}

Ultra-stripped SNe have been related to faint and fast optical transients, especially Type~Ib anc Ic SNe \citep{tauris2013,moriya2017}.  Some observed ultra-stripped SNe have early interaction signatures within a few days after the explosion \citep[e.g.,][]{de2018}.  These early interaction signatures have been connected to possible mass ejection shortly before SNe, but the estimated CSM mass has been of the order of $0.01~\Msun$ \citep[e.g.,][]{de2018}.  As shown in our ultra-stripped SN progenitor model, ultra-stripped SNe can have much more massive CSM of the order of $0.1~\Msun$.  The interaction signatures dominate in such cases and they are likely observed as interaction-powered SNe.  The SNe are expected to be of Type~Ibn SNe with observable interaction signatures of the helium-rich CSM \citep[e.g.,][]{pastorello2008}. 

\begin{figure}
 \begin{center}
  \includegraphics[width=8cm]{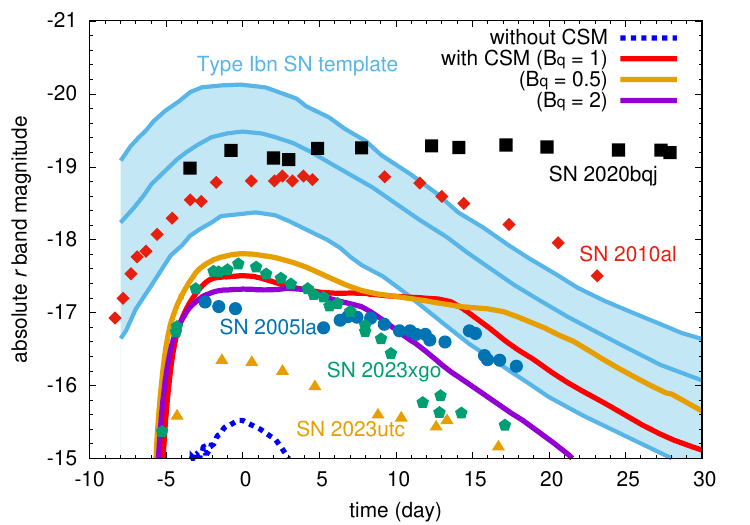}
 \end{center}
\caption{
Synthetic light curves of ultra-stripped SNe with and without CSM compared with Type~Ibn light curves in the \textit{r} band.  The Type~Ibn light-curve template is from \citet{hosseinzadeh2017}.  We also show the light curves of five Type~Ibn SNe not following the template behavior for comparison.  The circle symbols show the light curve of SN~2005la from \citet{pastorello200805la}.  The square symbols present the light curve of SN~2020bqj \citep{kool2021}.  The diamond symbols show the light curve of SN~2010al \citep{pastorello2015}. The pentagon symbols show the light curve of SN~2023xgo \citep{gangopadhyay2025}.  The triangle symbols show the light curve of SN~2023utc \citep{wang2025}.
 {Alt text: Lines show luminosity evolution and dots show observed light curves.  The x axis is the time from 0 to 30~days.  The y axis is the absolute r-band magnitude ranging from -15 to -21. } 
}\label{fig:rband}
\end{figure}

Figure~\ref{fig:rband} compares our synthetic ultra-stripped SN light curves with Type~Ibn SN light curves in the \textit{r} band.  A large fraction of Type~Ibn SNe show similar shapes with a rise and fall, and \citet{hosseinzadeh2017} provided a region where many Type~Ibn SN light curves are located as presented in Figure~\ref{fig:rband}.  We find that our synthetic light curves with the dense CSM are fainter than the major Type~Ibn luminosity region.  Our synthetic light curve has a short plateau-like phase after the peak during which the brightness does not change much.  Although the Type~Ibn SN template does not show such a plateau, some Type~Ibn SNe are known to have a plateau phase in their light curves.  We show some representative cases in Figure~\ref{fig:rband}.  Among them, we find that SN~2005la \citep{pastorello200805la,pastorello2015} has similar brightness and light-curve shape to our synthetic light curve.  We note, however, that SN~2005la shows weak hydrogen features and that this fact makes the scenario that SN~2005la is an ultra-stripped SN unfavorable.  This is because ultra-stripped SN progenitor systems are expected to lose their hydrogen during the common-envelope phase.  Moreover, any residual hydrogen is lost early in the subsequent Case~BB RLO prior to core collapse.  Thus, ultra-stripped SNe are not likely to show any hydrogen features.

\citet{gangopadhyay2025} recently suggested that SN~2023xgo is a potential Type~Ibn SN originating from an ultra-stripped SN progenitor.  We found that the luminosity evolution is indeed similar to our synthetic light curve, making SN~2023xgo a promising candidate for an ultra-stripped SN affected by the strong CSM interaction.  \citeauthor{gangopadhyay2025} estimated the SN ejecta and CSM properties in two methods assuming it is powered by the CSM interaction. One is based on the semi-analytic method formulated by \citet{chatzopoulos2012} and implemented in \texttt{REDBACK} \citep{sarin2024}.  This method obtained the ejecta mass estimate of $0.1~\Msun$ and the CSM mass of $0.4~\Msun$, which are similar to our initial conditions. The ejecta mass of $0.1~\Msun$, however, matches the lowest boundary in their ejecta mass range, indicating the difficulty in estimating the ejecta properties in interaction-powered SNe.  The other method is the non-equipartition model by \citet{maeda2022}.  This model resulted in a similar CSM mass estimate of $0.2~\Msun$, but the CSM density structure is estimated to be $\propto r^{-2.5}$, which is different from the CSM density structure in our model ($\propto r^{-1.3}$, Figure~\ref{fig:density}).  We emphasize that \citet{gangopadhyay2025} assumed the ejecta mass of $2~\Msun$ when applying the model of \citet{maeda2022}, which may have caused the difference.

In this work, we only investigated one particular ultra-stripped SN progenitor.  Ultra-stripped SN progenitors can have diverse ejecta mass and explosion energy \citep{suwa2015,muller2018,muller2019}.  The mass loss induced by the
violent silicon burning can occur in other ultra-stripped SN progenitors in a diverse way.  More systematic investigations of potential mass loss and subsequent explosions in ultra-stripped SN progenitors are required to identify Type~Ibn SNe from ultra-stripped SNe.  Encouragingly, some Type~Ibn SNe could indeed be associated with low-mass progenitors \citep[e.g.,][]{sanders2013,hosseinzadeh2019,wang2024} even though Type~Ibn SNe are often associated with massive Wolf-Rayet stars \citep[e.g.,][]{moriya2016,maeda2022}.  Literature estimates of the Type~Ibn SN event rate are less than 1~\% of core-collapse SNe rate \citep[e.g.,][]{ho2023,toshikage2024}, whereas the ultra-stripped SN event rate estimates are up to 1~\% of core-collapse SNe rate \citep[e.g.,][]{tauris2013}.  Thus, it is also possible that a large fraction of the ultra-stripped SNe are observed as Type~Ibn SNe.

It is generally difficult to estimate the nature of exploding stars under dense CSM in interaction-powered SNe because their observational signatures are dominated by the CSM interaction. Thus, determining if an observed Type~Ibn SN originate from an ultra-stripped SN progenitor with a dense CSM is a challenge. A small total radiated energy is one clue because ultra-stripped SNe are expected to have low explosion energies. In addition, if we can observe Type~Ibn SNe for more than around 150~days, the luminosity evolution starts to be set by the \Co decay (Figure~\ref{fig:bolometric}). The late-phase luminosity evolution allows us to estimate the \Ni mass in the ejecta, which can also provide a clue to judge if they originate from ultra-stripped SN progenitors. In addition,
ultra-stripped SN progenitors have a close compact companion star.  The compact companion star can also play a diverse role in determining the CSM properties and result in the formation of the dense CSM \citep[e.g.,][]{wu2022,wei2024,haynie2025,ko2025}.  In particular, if the ejected matter at 78~days before the core collapse accretes on to the companion compact star, the accretion may result in some brightening of the progenitor system even before core collapse. Such brightening may also explain precursors observed in some Type~Ibn SNe \citep{dong2024,brennan2024}.  It is also possible such brightening results in an X-ray flare.
Such a pre-SN signature can also allow us to judge if they originate from ultra-stripped SN progenitors.

\section{Conclusions}\label{sec:conclusions}

We have investigated the observational properties of an ultra-stripped SN model with a dense CSM formed by mass ejection 78~days before core collapse due to violent silicon burning. The dense CSM ($0.2~\Msun$) is more massive than the SN ejecta mass ($0.06~\Msun$), causing immediate deceleration of the SN ejecta and the release of a significant fraction of its kinetic energy of $9\times 10^{49}~\mathrm{erg}$ as radiation.  We expect that this interaction-powered SN is observable as a Type~Ibn SN because the dense CSM in this model is hydrogen-free and helium-rich.  Whereas our model results in a low-luminosity Type~Ibn SN, more investigations of ultra-stripped SNe with dense CSM are required to have an overall idea on the possible Type~Ibn SN properties that can be achieved by ultra-stripped SN progenitors.  Our work demonstrates that ultra-stripped SNe can end up with diverse SNe and further investigations of possible outcomes from ultra-stripped SNe are required to obtain a complete picture of ultra-stripped SNe.


\begin{ack}
We thank the anonymous referee for constructive comments that improved the manuscript.
Numerical computations were in part carried out on PC cluster at the Center for Computational Astrophysics, National Astronomical Observatory of Japan.
SB and MU are grateful to K. Nomoto for his hospitality at IPMU in the fall 2024.
Some of the underlying explosion simulations were performed on the Gadi supercomputer with the assistance of resources and services from the National Computational Infrastructure (NCI), which is supported by the Australian Government, and through support by an Australasian Leadership Computing Grant. Some of this work was performed on the OzSTAR national facility at Swinburne University of Technology. The OzSTAR program receives funding in part from the Astronomy National Collaborative Research Infrastructure Strategy (NCRIS) allocation provided by the Australian Government, and from the Victorian Higher Education State Investment Fund (VHESIF) provided by the Victorian Government.

\end{ack}

\section*{Funding}
This research is supported by the Australian Research Council (ARC) through the ARC's Discovery Projects (DP) funding scheme (DP240101786).
TJM is supported by the Grants-in-Aid for Scientific Research of the Japan Society for the Promotion of Science (JP24K00682, JP24H01824, JP21H04997, JP24H00002, JP24H00027, JP24K00668). 
M. Ushakova thanks the RSF for supporting the work on the development of radiation-hydrodynamics methods (grant no. 24-12-00141).
The work of ES was conducted under the state assignment of Lomonosov Moscow State University.
AH has been supported by a Research Award by the Alexander von Humboldt Foundation and by ARC DP240103174.

\section*{Data availability} 
The data underlying this article will be shared on reasonable request to the corresponding author.







\bibliographystyle{apj}
\bibliography{pasj}

\end{document}